# Implicit and Coupled Multi-Fluid Solver for Collisional Low-Temperature Plasma

Robert Arslanbekov[1] and Vladimir Kolobov[1,2]

[1]CFD Research Corporation, Huntsville, AL, USA

[2]University of Alabama in Huntsville, Huntsville, AL, USA

We present a new multi-fluid, multi-temperature plasma solver with adaptive Cartesian mesh (ACM) based on a full-Newton (non-linear, implicit) scheme for collisional low-temperature plasma. The particle transport is described using the drift-diffusion approximation for electrons and ions coupled to Poisson equation for the electric field. Besides, the electron-energy transport equation is solved to account for electron thermal conductivity, Joule heating, and energy loss of electrons in collisions with neutral species. The rate of electron-induced ionization is a function of electron temperature and could also depend on electron density (important for plasma stratification). The ion and gas temperature are kept constant. The spatial discretization of the transport equations uses a non-isothermal Scharfetter-Gummel scheme from semiconductor physics adapted for multi-dimensional ACM framework. We demonstrate the new solver for simulations of direct current (DC) and radio frequency (RF) discharges. The implicit treatment of the coupled equations allows using large time steps, and the full-Newton method enables fast non-linear convergence at each time step, offering greatly improved efficiency of fluid plasma simulations. We discuss the selection of time steps for solving different plasma problems. The new solver enables us to solve several problems we could not solve before with existing software: two- and three-dimensional structures of the entire DC discharges including cathode and anode regions with electric field reversals, normal cathode spot and anode ring, plasma stratification in diffuse and constricted DC discharges, and standing striations in RF discharges.

## 1 Introduction

Plasmas are characterized by a disparity of time scales and non-linear behavior. The disparity of the time scales comes from the large difference between the electron and ion mass. The nonlinearities appear from coupling charged particle transport with electric fields and from ionization processes, which are highly sensitive to electron energy spectra. The disparity of time scales can be addressed with implicit solvers.[1] Such solvers enable using time steps larger than the characteristic time scales of the fast processes by effectively removing the time derivatives from the corresponding transport equations for electrons. Ideally, implicit solvers allow obtaining steady-state solutions in a single time step by using non-linear iterations to address coupling and nonlinearities of equations describing plasmas.

These strategies have been implemented in some form in most of the existing kinetic and fluid plasma solves.[2] Existing multi-fluid, multi-temperature models use Finite Volume [3,4] or Finite Element [5] space discretization and solve plasma transport equations sequentially using iterations for solving each equation. The sequential solution reduces computer memory usage, but limits the time step by the requirement of convergence of the solution - the "sequentialization penalty". For thermal plasma, a fully implicit multi-fluid reactive model with a single temperature for all species was implemented [6] assuming quasi-neutrality and a local thermodynamic equilibrium model for electrons. Fully implicit solvers require large memory, which limits the size of the problem they can solve.

For kinetic plasma simulations, Particle-in-Cell (PIC) methods and discrete velocity models have been developed. They are very computationally expensive compared to the fluid models because they typically operate on the shortest time- and length scales and calculate the velocity distribution functions of plasma species rather than their macro-parameters such as density, mean velocity, and temperature. Implicit and semi-implicit particle-based and mesh-based kinetic solvers have also been developed.[7,8] It appears that the best procedure for addressing the disparity of time scales would be to separate the fast and slow processes in separate blocks that would enable individual control of the time steps and use "recycling". Such a "recycling" procedure is well known for the PIC methods.[9]

Recent demands for understanding and addressing the disparity of temporal, spatial, and energy scales in plasmas come from the development of adaptive kinetic-fluid solvers.[10] Many plasma problems require space, time, and model adaptation for an efficient solution.[11,12] The need for implicit solvers and adaptive time steps for solving such problems has been recognized.[13] Some problems require resolving both electron and ion time scales. A typical example is a low-temperature plasma maintained by alternating electric fields at high frequencies (from radio-frequency (RF) to microwave and optical range). Typically, at high frequencies ions do not respond to the time variations of the fields maintaining the plasma, they only respond to a slow-varying electric field generated by the plasma. In contrast, electrons could respond to the high-frequency field dynamics forming sheaths and skin layers, where electrons acquire kinetic energy for gas ionization. How to solve such problems in the general case with implicit coupled solvers? How to select the appropriate time step? Can we develop optimal strategies for simulations of collisional plasma operating at different frequency ranges? The present paper attempts to address some of these challenges.

Implicit methods or, more generally, Full Newton Methods (FNM) have been introduced in the late 80s and early 90s for modeling semiconductor devices.[14,15,16,17,18,19,20] They became a standard in most of the commercially available simulators, such as Medici, Dessis, and others. The NanoTCAD software [21,22] relies on FNM implementation on a binary Cartesian mesh. The FNM approach is also an essential part of many Computational Fluid Dynamics (CFD) codes[23,24,25,26] where it is used because of the numerical stiffness intrinsically present for many CFD problems. The FNM approach is typically applied to the set of Navier-Stokes equations for the density, momentum, and energy equations describing the mean flow properties while solving the often-stiff chemistry for species fractions in the same global matrix constructed from the Jacobians.

Automatic mesh generation and dynamic mesh adaptation to a solution and/or a changing geometry is a hot topic in computational physics. Adaptive Cartesian mesh (ACM) becomes popular in CFD and multi-physics simulations and is an essential part of today's CFD codes such as Converge,[27] Simerics,[28] FloEFD,[29] and others. We have implemented the ACM technique in our Adaptive Mesh and Algorithm Refinement (AMAR) framework,[30] which allows not only to adapt the mesh to locally required resolution but also to select kinetic or fluid models on a cell-by-cell basis. In the AMAR framework, an open-source Gerris flow solver, GFS,[31] is used for generating ACM around embedded boundaries. GFS produces a 2:1 balanced grid, which means that a) the levels of direct neighbors cannot differ by more than one and b) the levels of diagonal neighbors cannot differ by more than one. These additional constraints simplify the gradient and flux calculations. Fully threaded tree, the pointer-based data structure of GFS allows fast access to cell neighbors, which facilitates efficient implementation of dynamic AMR and parallelization. In addition to the cut-cell technique for the boundary treatment, we have also implemented the immersed boundary method for embedded boundaries.[32]

The present work is devoted to coupling adaptive Cartesian mesh with a full Newton solver for fluid plasma simulations. The authors are not aware of any publications describing the FNM implementation on the dynamically adaptive Cartesian mesh. For related work, we can mention a recently developed Zapdos code[33] in the MOOSE framework, which relies on the Finite Element Method (FEM) for solving PDEs. Zapdos implements the FNM via interfacing with PETSc[34] to solve the fluid plasma equations using a logarithmic transformation for the particle densities and the electron energy density to ensure the positivity of the solution. The log transformation introduces additional non-linearity and makes the implementation more involved compared to the natural variables used in the present work. Most of the current plasma codes for modeling gas discharge systems are based on a segregated approach. For example, our previous AMR plasma code[30] uses explicit solvers which are suitable only for solving problems at the fast (electron) time scales. The CFD-ACE-Plasma code[48] is based on implicit solvers, but the PDEs are solved sequentially (i.e., in uncoupled or loosely coupled manner; e.g., via additional sub-iterations), which may pose numerical issues for solving problems with strong coupling (e.g., large plasma densities).

The structure of the paper is as follows. Section 2 describes the basic equations of the multi-fluid, multi-temperature plasma model. Section 3 discusses in detail the implementation of the FNM-ACM solver for these equations and the selection of time steps for solving specific plasma problems. Section 4 contains simulation results and compares the efficiency of the explicit and FNM plasma solvers. Finally, Section 5 contains a conclusion and outlook.

# 2 Plasma Model Equations

The plasma model equations adopted in this work include the balance equations for the electron ($n_e$) and ion ($n_i$) densities coupled to Poisson equation for the electrostatic field:

$$q_e \frac{\partial n_e}{\partial t} - \nabla \cdot \boldsymbol{J}_e = q_e S_e, \tag{1}$$

$$q_e \frac{\partial n_i}{\partial t} + \nabla \cdot \boldsymbol{J}_i = q_e S_i, \tag{2}$$

In these equations $q_e = |e|$ is the absolute value of the electron charge, $S_k$ are the source (e.g., ionization) and sink (e.g., recombination) terms, $\boldsymbol{J}_e$ and $\boldsymbol{J}_i$ are the electron and ion current densities, respectively. The electron energy transport equation is:

$$\frac{\partial(n_e \varepsilon_e)}{\partial t} + \nabla \cdot \boldsymbol{\xi}_e = \boldsymbol{E} \cdot \boldsymbol{J}_e - n_e \sum_{r=1}^{N_{reactions}} K_r \Delta \varepsilon_r, \tag{3}$$

where $\varepsilon_e$ is the average electron energy, $K_r$ is the reaction rate constant and $\Delta \varepsilon_r$ is the electron energy change per electron per collision. For inelastic collisions $\Delta \varepsilon_r$ corresponds to the energy threshold of an $r$-type collision (e.g., excitation or ionization processes). The energy loss corresponding to elastic collisions is given by

$$\Delta \varepsilon_r = \frac{3}{2} \frac{m_e}{M_r} k_B (T_e - T_g), \tag{4}$$

where $m_e$ is the electron mass and $M_r$ is the mass of the heavy particle (e.g., atom) and $T_g$ is the common temperature of neutral plasma species.

The electron current density is

$$J_e = q_e D_e \nabla n_e - q_e n_e \mu_e \nabla \left(\varphi - \frac{k_B T_e}{q_e}\right), \tag{5}$$

or, in a different form

$$J_e = q_e D_e \nabla n_e + q_e n_e \left\{\mu_e \nabla(-\varphi) + D_e \frac{\nabla T_e}{T_e}\right\}. \tag{6}$$

Similarly, for the ion species ($i = i_1, i_2, ...$)

$$J_i = -q_e Z_i D_i \nabla n_i + q_e Z_i n_i \left\{\mu_i \nabla(-\varphi) - D_i \frac{\nabla T_i}{T_i}\right\}. \tag{7}$$

Here, the electron and ion mobilities are $\mu_e$ and $\mu_i$, and the corresponding diffusion coefficients, $D_{e,i} = \frac{k_B T_{e,i}}{q_e} \mu_{e,i}$. The drift-diffusion (DD) approximation for the electron and ion fluxes neglects inertial effects, which is a reasonable approximation for collision-dominated gas discharge plasmas. Furthermore, the ion motion is dominated by drift. The ion diffusion term is small in most of the computational domain, and can only be comparable to the drift term at the points of zero electric field. It is known that the ion inertia effects are important in weakly collisional plasma, and they will be added to our solver in future work.

The electron energy flux $\xi_e$ in Eq. (3) can then be written as:

$$\xi_e = -\kappa_e \nabla T_e - (\varepsilon_e + k_B T_e) \frac{J_e}{q_e}. \tag{8}$$

The average electron energy $\varepsilon_e$ consists of thermal energy, $3/2 k_B T_e$, and the electron kinetic energy $1/2 m_e v^2$. We can neglect the electron directed kinetic energy compared with the thermal energy to obtain:

$$\xi_e = -\kappa_e \nabla T_e - \frac{5 k_B T_e}{2} \frac{J_e}{q_e}. \tag{9}$$

According to (9), the electron energy flux $\xi_e$ contains thermal conduction and convection terms. The thermal conductivity coefficient ($\kappa_e$) for electrons is given by the Wiedemann-Frantz law:

$$\kappa_e = \left(\frac{5}{2} + c_e\right) \frac{k_B^2}{q_e} T_e \mu_e n_e. \tag{10}$$

For semiconductors, the value of $c_e = -0.5$ is commonly used,[18] while $c_e = 0$ is common for plasma. We will illustrate below how different values of $c_e$ can affect plasma stratification. In most publications devoted to plasmas, the energy density flux is used in the form:

$$\xi_e = -\frac{5}{3}D_e \nabla(n_e \varepsilon_e) - \frac{5}{3}\mu_e \nabla \varphi (n_e \varepsilon_e). \tag{11}$$

However, the energy flux expressed by Eq. (9) has a more transparent physical meaning since it separates the thermo-diffusion term explicitly. Eq. (9) also indicates that in the quasi-neutral plasma, where the electron flux ($J_e$) transforms into a ambipolar diffusion flux, $J_a \sim J_i$, the thermal diffusion (proportional to the free electron diffusion coefficient) dominates over convection. Indeed, the ratio of the free electron diffusion coefficient to the ambipolar diffusion coefficient scales as $D_e/D_a \sim \sqrt{M/m_e} \gg 1$ for typical $T_e \gg T_i$.

The system of the particle density and electron energy transport equations is completed by the Poisson equation for the electrostatic potential $\varphi$:

$$\nabla \cdot (\epsilon \mathbf{E}) = \nabla \cdot (-\epsilon \nabla \varphi) = q_e \left( \sum Z_i n_i - n_e \right), \tag{12}$$

where $\epsilon = \epsilon_r \epsilon_0$ is the permittivity of the plasma and $\epsilon_0$ is that of vacuum (for gas, $\epsilon_r = 1$).

The system of plasma equations is fully defined by setting appropriate boundary conditions at the plasma device walls. At metal walls, the following boundary conditions are set for the electron and ion fluxes:

$$\mathbf{n} \cdot \mathbf{J}_e = q_e \left( \frac{1}{4} v_e n_e - \sum \gamma_i (\mathbf{n} \cdot \mathbf{\Gamma}_i) \right), \tag{13}$$

$$\mathbf{n} \cdot \mathbf{J}_i = q_e \left( \frac{1}{4} v_i n_i + a n_i \mu_i (\mathbf{n} \cdot \mathbf{E}) \right), \tag{14}$$

where $\gamma_i$ is the secondary electron emission coefficient due to $i^{th}$ ion bombardment, the ion flux $\mathbf{\Gamma}_i = \mathbf{J}_i/q_e$, and

$$a = \begin{cases} 1, & (\mathbf{n} \cdot \mathbf{E}) > 0 \\ 0, & (\mathbf{n} \cdot \mathbf{E}) \leq 0 \end{cases},$$

with $\mathbf{n}$ being a wall directed unit normal vector. For the energy flux, the following boundary condition is set:

$$\mathbf{n} \cdot \xi_e = \frac{1}{3} v_e (n_e \varepsilon_e) - 2k_B T_e \sum \gamma_i (\mathbf{n} \cdot \mathbf{\Gamma}_i). \tag{15}$$

At metal boundaries, the electrostatic potential can be set to given values (so-called Dirichlet-type boundary conditions), which are in turn explicitly prescribed or computed from external circuit solutions.

At dielectric walls, the same set of boundary conditions defined in Eqs. (13), (14), and (15) is used (with typically $\gamma_i = 0$). In order to compute the plasma potential at these boundaries, we introduce a surface charge, $\sigma$, for which we solve the following time dependent (local) balance equation:

$$\frac{\partial \sigma}{\partial t} = \sum Z_i (\boldsymbol{n} \cdot \boldsymbol{J}_i) - (\boldsymbol{n} \cdot \boldsymbol{J}_e) \tag{16}$$

at all boundary faces. The surface charge density is then used to compute the appropriate boundary conditions at boundary faces for the electrostatic potential through Gauss' law:

$$\epsilon(\boldsymbol{n} \cdot \boldsymbol{E}) = \epsilon\big(\boldsymbol{n} \cdot (-\nabla \varphi)\big) = \sigma. \tag{17}$$

## 3 Full Newton Method Implementation in the ACM Framework

### 3.1 Discretization of the Enhanced SG Scheme

To discretize the plasma equations, we have utilized the Finite Volume (FV) approach with a cell-centered definition for all solution variables. Figure 1 illustrates a control-volume (CV) cell together with neighboring cells of different refinement levels. In our ACM framework, the refinement level of neighbor cells cannot be larger than one, so that only one hanging node per any given face is allowed. The FV spatial reconstruction scheme illustrated in Figure 1 is based on the notion of bringing the neighbor cell values to the same level along the coordinate axes. This is done by invoking neighboring cells in the perpendicular direction to the current face of the CV cell across which the SG fluxes are computed. Then, the right-state (whose locations are marked by cross symbols in Figure 1) is reconstructed by a linear combination of all involved neighbor cell-centered values while the left-state consists of the cell-centered values of the CV cell itself. This way, 2nd order spatial accuracy is maintained when performing reconstruction at faces involving not axis aligned cells (so-called transversal or diagonal neighbors).[31]

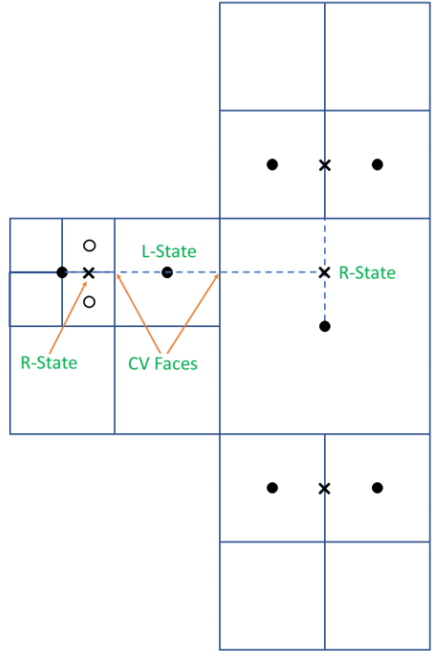

*Figure 1. Schematic representation of face reconstruction on ACM with the cell to the left being one-level up and the cell to the right being one-level down (after Ref. [31]). Cell faces (here, right, and left) are shown by arrows. Also shown are right-state locations (crosses) involved in face reconstruction in the developed SG scheme.*

The SG scheme for computing the particle and energy fluxes at cell faces is obtained by assuming that, except the charged species densities, all other quantities such as the electric field, electron and ion mobilities and their current densities remain unchanged between the left ($L$) and right ($R$) states (Figure 1) of the solution vector across any given face of a control volume cell.[14,16,18] Then, the electron and ion density fluxes at such faces are computed as

$$J_e^f = \frac{q_e \mu_e^f U_{T_e}^f}{\Delta_f}\left[n_{e,R} B\left(\frac{\varphi_R - \varphi_L}{U_{T_e}^f}\right) - n_{e,L} B\left(-\frac{\varphi_R - \varphi_L}{U_{T_e}^f}\right)\right], \qquad (18)$$

$$J_i^f = \frac{Z_i q_e \mu_i^f U_{T_i}^f}{\Delta_f}\left[n_{i,L} B\left(\frac{\varphi_R - \varphi_L}{U_{T_i}^f}\right) - n_{i,R} B\left(-\frac{\varphi_R - \varphi_L}{U_{T_i}^f}\right)\right], \qquad (19)$$

where the face quantities (locations indicated are arrows in Figure 1) are labeled by "$f$" superscript index, $U_{T_{e,i}} = k_B T_{e,i}/q_e$ is the thermal electrical constant, $\mu_{e,i}^f$ are electron and ion mobilities at cell faces, $\Delta_f$ is the face normal distance between left and right (neighbor) cell centers or positions (depending on the neighboring cell refinement levels), and $B(x) = x/(e^x - 1)$ is the Bernoulli function.

As described earlier, the left-state values of the solution vector represent those at the cell-center of a control-volume cell, while those from the right-state are reconstructed using a linear combination of values from the cell centers of neighboring cells thus providing the required 2$^{nd}$ order spatial accuracy during the reconstruction step.[31] From the above-defined scheme, one can see that to compute the electron and ion fluxes one needs to know face values of the electron and ion mobilities. The face values are interpolated as

$$\mu_{e,i}^f = \mu_{e,i}^L + w^f(\mu_{e,i}^L - \mu_{e,i}^R), \tag{20}$$

with a face weight, $w^f$, determined by the face neighbor geometry (fine-fine, coarse-fine, etc.; e.g., $w^f = 0.5$ on a fine-fine interface) while the right-state values are reconstructed by a linear combination of neighboring cell-center values (see Ref. [31]). Similarly, $U_{T_{e,i}}^f$ and all other face properties such as electron and ion diffusion coefficients are computed.

An important feature of the developed model is an extended, nonisothermal SG scheme. The FV variant of this extended SG scheme for the electron current continuity equation including a position-dependent variation of the electron temperature. The SG formulations in Eqs. (18) and (19) are valid under the assumption of spatially uniform electron temperature. When the electron temperature varies across neighboring cells, the nonisothermal SG scheme for the electron density flux becomes:[18,19]

$$J_e^f = \frac{q_e D_e^f (T_{e,R} - T_{e,L})}{\Delta_f \ln\left(\frac{T_{e,R}}{T_{e,L}}\right)} \left[B(\tilde{\Delta})\frac{n_{e,R}}{T_{e,R}} - B(-\tilde{\Delta})\frac{n_{e,L}}{T_{e,L}}\right], \tag{21}$$

where

$$\tilde{\Delta} = \frac{\ln(T_{e,R}/T_{e,L})}{(T_{e,R} - T_{e,L})}\left[\frac{q_e}{k_B}(\varphi_R - \varphi_L) - 2(T_{e,R} - T_{e,L})\right]. \tag{22}$$

We now describe our treatment of the electron energy transport equation. There are several approaches to the energy flow's discretization in the nonisothermal SG scheme.[20] The most consistent way is to solve the current equation together with the energy flow equation, but this method results in an infinite-series solution. Tang[14] obtained the electron density by separately solving for it from the current expression (21) and the energy flow equation, and obtained large differences between the electron densities from these two solutions. Forghieri et al.[16] assumed an exponential profile of the electron density for the discretization of the electron energy transport equation. Later, Choi et al.[18] proposed a discretization scheme, where the electron density was obtained from the current equation (21) and substituted into the energy flow equation. The resulting approach decreased the mismatch between the electron densities obtained from these equations, helped decrease the error of electron temperature and enhance the convergence.[18] We adopt Choi's discretization scheme, in which the electron energy flow is written as [20]

$$S_e^f = -\left(\frac{5}{2} + c_e\right)\frac{k_B D_e^f}{\Delta_f}\frac{(T_{e,R} - T_{e,L})}{\ln(T_{e,R}/T_{e,L})}\left[B(\tilde{\Delta})\frac{B(\Phi)}{B(\tilde{\Phi})}n_{e,R} - B(-\tilde{\Delta})\frac{B(-\Phi)}{B(-\tilde{\Phi})}n_{e,L}\right], \tag{23}$$

where

$$\widetilde{\Phi} = \frac{\ln(T_{e,R}/T_{e,L})}{(T_{e,R} - T_{e,L})} \left[ \frac{q_e}{k_B}(\varphi_R - \varphi_L) - (T_{e,R} - T_{e,L}) \right] - \ln\left(\frac{n_{e,R}}{n_{e,L}}\right), \quad (24)$$

with

$$\Phi = \frac{\frac{5}{2}}{\frac{5}{2} + c_e} \widetilde{\Phi}. \quad (25)$$

For plasma systems, $c_e = 0$, and $\Phi = \widetilde{\Phi}$.

Another critical point in the implemented FNM model is the treatment of the Joule (Ohmic) heating source, which often represents a source of numerical instabilities. In conventional (non-FNM) plasma codes, an implicit treatment of the electron energy source term, based on linearization for the electron mean energy has been suggested.[4] This approach made it possible to increase the time step by several orders of magnitude. In our FNM approach, we have attempted two techniques. In the first one, the Joule heating term was rewritten as[18]

$$\boldsymbol{E} \cdot \boldsymbol{J}_e = -\nabla \cdot (\varphi \boldsymbol{J}_e) + q_e \varphi \left( \frac{\partial n_e}{\partial t} + S_e \right). \quad (26)$$

In the second approach, by converting the CV cell ($\Omega$) integration to summation over cell faces, we obtained [17,20]

$$\int_\Omega \boldsymbol{J}_e \cdot \nabla \varphi dV = \int_\Omega \nabla \cdot (\varphi \boldsymbol{J}_e) dV - \int_\Omega \varphi \nabla \cdot (\boldsymbol{J}_e) dV = \sum_f (\varphi_f - \varphi_L) J_e^f \Delta S^f, \quad (27)$$

with $\Delta S^f$ being the face surface area and $f$ index denotes summation over cell faces. We have implemented and tested both these methods and found that the second scheme gives more robust results and better non-linear convergence at each time step.

The implicit treatment of the boundary conditions is essential for the FNM scheme consistency to achieve better convergence properties. The boundary conditions at metal electrodes and dielectric walls are treated implicitly by properly setting the corresponding fluxes across the boundary cell faces and by then computing the corresponding Jacobian terms.

In the present work, the source terms, $S_e$ and $S_i$, included electron-induced ionization and volume recombination. Generally, the ionization rate should be obtained from a solution of the electron Boltzmann equation: [35]

$$R_e = \frac{\langle v_{ion} \rangle}{N} = \sqrt{\frac{2e}{m_e}} \int_0^\infty u \sigma_{ion}(u) f(u) du. \quad (28)$$

where $f(u)$ is the electron energy distribution function (EEDF), $u$ is the electron kinetic energy, and $\sigma_{ion}(u)$ is the ionization cross-section. By solving the *local* Boltzmann equation for electrons over a range of $n_e$ and $T_e$, one can obtain look-up-tables for the ionization rate as a function of $n_e$ and $T_e$ [48,36].

Instead of using this procedure, in the present work, the electron-induced ionization rate was expressed in

the Arrhenius form:

$$R_e(T_e) = AT_e^B e^{-E_a/T_e}, \qquad (29)$$

where $A$, $B$, and $E_a$ are constants, and $S_e = S_i = n_e R_e$. For modeling plasma stratification (see below), we also introduced a non-linearity arising due to Maxwellization of the EEDF:[37,38]

$$R_e(n_e, T_e) = AT_e^B e^{-E_a/T_e} \begin{cases} \exp\left(\frac{n_e}{n_c}\right), & n_e < n_1 \\ \exp\left(\frac{n_1}{n_c}\right), & n_e > n_1 \end{cases}. \qquad (30)$$

Here, $n_c$ controls the rate of the non-linear dependence, and $n_1$ defines the saturation value. The volume recombination was expressed in the form:

$$S_{recomb} = \beta n_i n_e. \qquad (31)$$

The recombination rate, $\beta$, was varied to study the transition between diffuse to constricted discharges. By gradually increasing $\beta$, we have observed a transition from diffuse to constricted discharge and analyzed the properties of striations in the diffuse and constricted column.[37] In our simulations reported below, the source terms were treated explicitly both for electrons and ions. For our purposes, this simple treatment of the source terms was sufficient.

The mesh refinement/coarsening criteria can be specified in the simulation scripts based on a combination of several events (see Ref. [31] for details). In these events, the minimum and maximum refinement levels are specified together with a set of sensitivity thresholds (or minimum variation of solution variable to be resolved across all neighboring cells of any given cell) for each particular grid adaptation event, which can be a function of cell position and solution variables. Such criteria can be based on computational cell locations (e.g., proximity to the plasma walls and electrodes), gradients or magnitudes of the solution variables, or any functions or combinations of such variables. Hence, depending on the problems being solved, one can build a quite sophisticated set of grid adaptation events to ensure proper grid resolution and solver efficiency for both steady-state and transient (dynamic grid adaptation) problems. In this work, we have used a simple set of grid adaptation events based on gradients of the electrostatic potential and electron density. This was sufficient for plasmas controlled by ambipolar diffusion. More sophisticated grid adaptation criteria have been previously used for simulations of high-pressure gas breakdown and streamer simulations with AMR.[39]

The grid adaptation events can specify any given time step interval (in terms of time step number or physical time) at which these events are executed. For steady-state simulations, we typically adapt the grid every 100-1000 time steps, while for transient simulations, the grid is adapted more often (e.g., every time step or every 10[th] step).

We paid attention to properly redistributing conserved quantities across smaller cells upon refinement. The zeroth-order algorithm would just uniformly distribute the parent cell solution across all children cells of the next refinement level. The reconstruction approach we are using is based on the first-order spatial accuracy approximation which employs slope-limited (e.g., Van Leer) gradients. The gradients are computed at cell centers of parent (coarse) cells, from which the values at cell centers of children (fine) cells are constructed. For some variables (such as electron density), to ensure further robustness, we apply logarithmic transformation (which preserves positiveness). During cell coarsening, the solution variables in the children cells are cast into their larger parent cell using a cell volume averaging technique, which ensures proper quantity conservation.

## 3.2 FNM Implementation

In the section, we describe the implementation of the FNM technique for solving the discretized system of the plasma transport equations coupled to the Poisson equation. In the FNM method, the full set of plasma equations with $N$ ion species is represented as

$$\boldsymbol{F}(\varphi, n_e, n_{i1}, \ldots, n_{iN}, T_e) = 0, \tag{32}$$

It is also possible to use the total electron energy ($\mathcal{E}_e = n_e \bar{\varepsilon}_e$) as an independent variable (instead of the electron temperature). Such an approach has been used in Ref. [20] to secure Jacobian matrix diagonal dominance (especially in the regions with low electron density) as well as in Ref. [40] in combination with log transformation of $\mathcal{E}_e$. We have implemented both approaches in our FNM scheme, and for the conducted test problems have not observed significant advantages of the latter approach. However, as we continue implementing new features (such as implicit treatment of the finite-rate chemistry) and broaden the range of test problems, we keep assessing these two approaches for better performance.

Then, in the selected FNM scheme, the complete set of multi-ion plasma equations with $N_{ions}$ ion species can be represented in a vector form as

$$\boldsymbol{F}(\boldsymbol{Q}) = 0, \tag{33}$$

where the $N_{ions} + 3$ solution vector (defined at cell centers of control-volume cells) is

$$\boldsymbol{Q} = \begin{pmatrix} \varphi \\ n_e \\ n_{i1} \\ \ldots \\ n_{iN} \\ T_e \end{pmatrix}, \tag{34}$$

$$\boldsymbol{F} = \begin{pmatrix} F_\varphi \\ F_{n_e} \\ F_{n_{i1}} \\ \ldots \\ F_{n_{iN}} \\ F_{T_e} \end{pmatrix} = \begin{pmatrix} \nabla \cdot (-\epsilon \nabla \varphi) - q_e \left( \sum Z_i n_i - n_e \right) \\ q_e \dfrac{\partial n_e}{\partial t} - \nabla \cdot \boldsymbol{J}_e - q_e S_e, \\ q_e \dfrac{\partial n_{i1}}{\partial t} + \nabla \cdot \boldsymbol{J}_{i1} - q_e S_{i1} \\ \ldots \\ q_e \dfrac{\partial n_{iN}}{\partial t} + \nabla \cdot \boldsymbol{J}_{iN} - q_e S_{iN} \\ \dfrac{\partial (n_e \frac{3}{2} k_B T_e)}{\partial t} + \nabla \cdot \boldsymbol{\xi}_e - \boldsymbol{E} \cdot \boldsymbol{J}_e + n_e \sum_r K_r \varepsilon_r \end{pmatrix}. \tag{35}$$

As one can see, we include the ion species into the same coupled matrix together with electrons and electrostatic potential. This was done by following the techniques used typically in semiconductor device modeling, where, as we recall, the difference between the hole and electron mobilities is not as large as in plasmas. For plasma systems, it may be advantageous to separate (or segregate) the electron and ion transport equations, thus considerably reducing the matrix size and the computational cost. In this work, we proceed however with the fully coupled approach to illustrate the method capabilities. Such a fully coupled method may be advantageous for modeling ion-ion plasmas (where the negative electric charge is dominated by the negative ions while the electric power is still absorbed by the electrons) and magnetized plasmas where electron transport is strongly affected by the magnetic field.

By introducing the following update rule (for each implicit time step advance $m \leftarrow m + 1$) with $k$ being the non-linear (implicit) sub-iteration index, one can write down a compact form with an update vector $\Delta \boldsymbol{Q}$

$$\boldsymbol{Q}^{m+1,k+1} = \boldsymbol{Q}^{m+1,k} + \Delta \boldsymbol{Q}^{m+1,k+1}, \tag{36}$$

where the initial condition for each Newton sub-iteration $k$ is:

$$\boldsymbol{Q}^{m+1,k=0} = \boldsymbol{Q}^m. \tag{37}$$

Following Ref. [18], we employ the first-order backward-differentiation-formula to the time-derivative terms. For time discretization, composite techniques employing both the trapezoidal rule and the second-order backward-differentiation-formula was used, as described in Ref. [18]. A Newton iteration is then solved as

$$\frac{\partial \boldsymbol{F}(\boldsymbol{Q}^k)}{\partial \boldsymbol{Q}} \Delta \boldsymbol{Q}^{k+1} = -\boldsymbol{F}(\boldsymbol{Q}^k), \tag{38}$$

where the time marching index $m$ is omitted for brevity.

In the FV formulation, the discretized flux vector $\boldsymbol{\Phi}$ across each face of a control-volume cell becomes a function of the left state $\boldsymbol{Q}_L$ (composed of the cell center values of the control-volume cell, i.e., $\boldsymbol{Q}_L = \boldsymbol{Q}$) and the right state $\boldsymbol{Q}_R$ (involving direct as well as indirect neighboring cell values, see Figure 1), i.e., $\boldsymbol{\Phi} = \boldsymbol{\Phi}(\boldsymbol{Q}_L, \boldsymbol{Q}_R)$. Then the discretized form of the Jacobian matrix for a Newton iteration takes two forms for the left and right (neighbor) state vectors. For plasma with one ion-species, we can write:

$$\mathbf{\Upsilon}_L^k = \frac{\partial \mathbf{F}(\mathbf{Q}^k)}{\partial \mathbf{Q}_L} = \begin{pmatrix} \frac{\partial F_\varphi}{\partial \psi_L} & \frac{\partial F_\varphi}{\partial n_{e,L}} & \frac{\partial F_\varphi}{\partial n_{i,L}} & \frac{\partial F_\varphi}{\partial T_{e,L}} \\ \frac{\partial F_{n_e}}{\partial \varphi_L} & \frac{\partial F_{n_e}}{\partial n_{e,L}} & \frac{\partial F_{n_e}}{\partial n_{i,L}} & \frac{\partial F_{n_e}}{\partial T_{e,L}} \\ \frac{\partial F_{n_i}}{\partial \varphi_L} & \frac{\partial F_{n_i}}{\partial n_{e,L}} & \frac{\partial F_{n_i}}{\partial n_{i,L}} & \frac{\partial F_{n_i}}{\partial T_{e,L}} \\ \frac{\partial F_{T_e}}{\partial \varphi_L} & \frac{\partial F_{T_e}}{\partial n_{e,L}} & \frac{\partial F_{T_e}}{\partial n_{i,L}} & \frac{\partial F_{T_e}}{\partial T_{e,L}} \end{pmatrix}, \tag{39}$$

where some of the Jacobian matrix elements are zero (such as, $\frac{\partial F_\varphi}{\partial T_{e,L}}$, and $\frac{\partial F_{T_e}}{\partial n_{i,L}}$). Similarly, the Jacobian matrix $\mathbf{\Upsilon}_R^k$ corresponding the right state vector $\mathbf{Q}_R$ can be written down. In the implemented numerical scheme, all Jacobian matrix entries are computed analytically by using free online differentiation tools based on the density and energy flux definitions given above. This makes the code run faster and more robust, as no numerical differentiation needs to be performed.

The Jacobian matrix of the fully discrete system is assembled as a sparse matrix using the compressed row storage format. The elements of this matrix are found using analytically computed derivatives of the SG fluxes presented in the paper using discrete values of the solution vector represented at each face of the control-volume. The Jacobian matrix is allocated at the beginning of each time step based on the current grid state together with the cell-connectivity matrix. The cell-connectivity matrix is computed only once at the beginning of each time step. The elements of the Jacobian matrix are updated at each non-linear Newton sub-iteration. At the end of the time step both matrices are deallocated. Although these were not very time-consuming operations for the problems solved in the present paper, the matrices' allocation and the computation of cell-connectivity need to be done only after the computational grid has undergone adaption. This approach will be implemented in future work to increase the code's efficiency for larger problems. Such an implementation is straightforward in the current framework.

### 3.3 Linear Matrix Solver

At the core of the implemented FNM is a linear matrix solver which solves the resulting system of linear equations:

$$\mathbf{\Upsilon}_{L,R}^k \Delta \mathbf{Q}_{L,R}^{k+1} = -\mathbf{F}(\mathbf{Q}^k). \tag{40}$$

The corresponding matrix is of general unsymmetrical, sparse type. As pointed out in Ref. [41], linear systems arising from the discretization of the TCAD (and thus plasma) models can be highly ill-conditioned and therefore quite challenging for direct and preconditioned iterative solvers. That work then discusses recent advances in the development of robust direct and iterative sparse linear solvers. It was obtained in Ref. [41] that for the preconditioned iterative Krylov subspace solvers, nonsymmetric permutations combined with scaling unsymmetrical reorderings gave the best results in terms of the number of required iterations and the time to compute the solution. As it is further discussed in detail in Ref. [20], there are 3 main ways to solve this equation: the direct method (LU decomposition), fixed-point iteration algorithms

(such as Gauss-Seidel and super relax iteration schemes), and Krylov subspace iteration methods (conjugate gradients class, minimum residual class, and others). With the direct methods being computationally costly (matrix bandwidths of ~15-30 are obtained in our typical TCAD and plasma simulations), one tends to rely on the iteration methods which in turn must tackle two issues: large matrix condition numbers (due to which the LU algorithm can fail and the number of iterations in Krylov subspace iteration methods drastically increase) and the round-off errors (important to minority carrier densities in semiconductors and sheath regions in plasmas). Algorithms relying on Transport Free Quasi-Minimal Residual (TFQMR) and Generalized Minimal Residual (GMRES) methods, as well Conjugate Gradient Squared method (CGS), Bi-Conjugate Gradient (BICG), Bi-Conjugate Gradient Stabilized (BCGS) were observed to give good results but they must be combined with ways of minimizing the residual, see Ref. [20] for details. The current (at the time of writing Ref. [20] ) commercial TCAD software packages, such as Medici integrated the LU and CGS methods, with Dessis adopting the LU and TFQMR methods, while the TCAD code in Ref. [20] builds upon the LU method (for smaller problems) as well as on the Krylov subspace methods available in PETSc.[42] The modern commercial TCAD software packages have been reported to successfully make use of such matrix solvers as PARDISO.[43]

In the present work, we used a high-performance iterative solver [44] available in our NanoTCAD framework. It uses high order preconditioning by incomplete decomposition to ensure good accuracy, reliable stability, and fast convergence. The resulting linear algebraic system is solved using a CGS-type iterative method with preconditioning by incomplete decomposition. To avoid diagonal pivot degeneration, the Kershaw diagonal modification is used, and the apparent computational complexity of the solver estimates as $\sim O(N^{5/4})$, which allows efficient computations for larger matrices arising in multi-dimensional settings. Alternatively, to facilitate the usage of the developed techniques by other research groups, we are working on interfacing and adapting the AMR FNM framework to open-source linear matrix solvers available in the PETSc and Trilinos[45] suites. This will also allow us to have more comprehensive customization over the broad range of available solver types and their tuning parameters to achieve better convergence for each problem of interest, as well as to tackle code's efficient parallelization. The results of these developments will be reported elsewhere.

## 3.4 Non-Linear Convergence and Time Stepping

In the developed FNM approach, the system of governing equations is advanced from one time to the next, $t \to t + \Delta t$ (time marching index $m$ in Eq. (36)), while performing non-linear sub-iterations for each given time step (index $k$ in Eqs. (36) and (38)). Because of the nature of the transient governing equations, non-linear convergence within each time step is largely controlled by the selected time step ($\Delta t$), which defines the diagonal dominance of the Jacobian matrix. Also, because of the AMR capabilities, the convergence (both local and global) can be efficiently controlled by dynamically adapting the computational grid to resolve the critical plasma features, e.g., large space charge and strong electric field. The time local (within a given time step, index $k$) convergence of the FNM technique (as any Newton iteration technique) is also determined strongly by initial conditions (set at $k = 0$ in Eq. (37)). If these initial conditions (at time $t$) are close to the solution at the end of the sub-iteration cycle (at time $t + \Delta t$), when large time steps can be used. This is typically the case for steady-state (e.g., DC) discharges. For transient simulations of RF discharges or gas breakdown simulations, smaller time steps are required. Since the system of governing equations is strongly non-linear and depends greatly on the initial conditions, it is not possible to provide a general recipe for selecting the maximum allowed time step $\Delta t$ *a priory*, contrary to the explicit solvers where the time step can be most often estimated accurately from the CFL condition and from the dielectric relaxation time controlling the particle-field coupling. In our simulations with the FNM scheme, the time step was determined by experimenting with the non-linear convergence rates over each time step (typically, convergence over 4−5 orders of magnitude suffices). Since an implicit time step costs more than an explicit time advance, one can expect that the FNM scheme will be most efficient for steady-state or slowly-varying problems by allowing large time steps. At the same time, transient problems evolving on the (fast) electron-time scale can be more efficiently solved by explicit techniques.

In our typical FNM plasma simulations, low initial plasma density was set so that the applied electric field was only slightly disturbed by the plasma. Once a voltage was applied to one of the electrodes (at time $t = 0$), electron motion and multiplication started at the fast electron time scale. The electron motion produced space charge and high electric fields, enhancing further electron multiplication and development of electron avalanche. During this initial highly transient phase, the time steps $\Delta t$ were typically set to 0.05−0.1 ns for the problems described below. Such small initial time steps originate from rapidly evolving plasma dynamics during these times. After completing this stage, the plasma dynamics occur on the slow ion time scale, and the FNM scheme allows increasing the time steps. In our simulations, we have found that and it is advantageous to gradually ramp up $\Delta t$ during the following few 1000's time steps with the final allowed time step being set to the values which ensure good non-linear convergence (typically, 4−5 orders of magnitude). Typical achieved $\Delta t$'s at which our developed FNM ranged between a few ns for RF plasma problems and 200−500 ns for DC problems. A typical example of a setup involving a 2D DC plasma cell is shown in Figure 2. The simulation was executed in Ar gas at 400 mTorr, with the left electrode being a grounded cathode ($\gamma = 0.1$) and the right electrode being an anode (applied voltage 200 V), the top and bottom boundaries being dielectric walls, and inter-electrode gap of 2 cm. AMR was based on magnitudes of space charge and electric field. A fixed number of non-linear sub-iterations (10) were performed with the total number of time steps of 15000. During the first 5000 steps, $\Delta t$ was increased to 200 ns. The time histories of residuals plotted in Figure 2 show that for the three main solution variables (electrostatic potential, electron density, and temperature), the local residuals (index $k$ in Eqs. (36) and (38)) drop by a least 4 orders of magnitude (by almost 10 orders of magnitude at the beginning of simulation when the time step is still small), while the global residuals (index $m$ in Eq. (36)) dropped by almost 12 orders of magnitude thus indicating complete convergence. Such convergence rates (to almost machine precision) may not be required for most of the typical plasma simulations, but they demonstrate superior capabilities of the developed FNM code. Our preliminary results indicate that DC plasma simulations with the developed FNM code are up to a factor of 100 more efficient than those obtained with our explicit code.[30] More detailed comparison studies are in order when fully implicit treatment of finite-rate chemistry is implemented, which will allow increasing further the efficiency of the new FNM framework for a broad range of plasma problems.

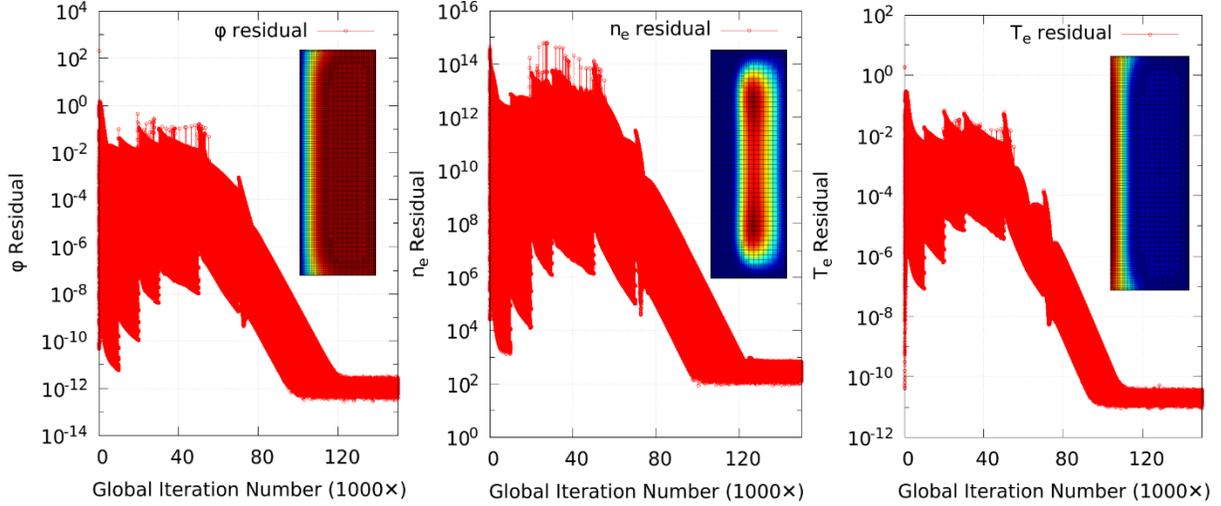

*Figure 2. Example of residual evolution obtained with the implemented FNM for short 2D DC discharge cell (Ar, pressure 400 mTorr, gap 2 cm, applied DC voltage 200 V) with AMR and final time step, Δt, of 200 ns. Shown are residuals for electrostatic potential, electron density, and temperature vs global iteration number). The inserts show corresponding plasma profiles with overlaid AMR grid.*

# 4 Results of Simulations

We have applied the new solver to simulations of various DC and RF discharges in 2D and 3D settings. The code is capable of handling 2D-axisymmetric geometries, which are typical for these systems, by properly setting the face metrics in the electron and ion particle and the energy fluxes across faces of control-volume cells.

## 4.1 Direct Current Glow Discharges

Figure 3 shows the results of 2D simulations of a DC discharge in a long cylindrical dielectric tube of radius $R = 1$ cm and the inter-electrode length $d = 7$ cm. The cathode is grounded and a voltage of 235 V is applied to the anode. The initial plasma density was set to ~$10^8$ cm$^{-3}$. The time step for this case was 50 ns.

The results shown in Figure 3 illustrate the typical structure of a DC glow discharge containing a cathode region, an axially uniform positive column, and an anode region. The plasma density (a) has a large peak in a negative glow near the cathode, passes through a minimum in the Faraday dark space before reaching an axially constant value in positive column plasma, and decays near the anode. The electric potential distribution (b) corresponds to connecting the equipotential electrodes with a positive column plasma where the radial (ambipolar) electric fields are established to equalize the fluxes of electrons and ions to the wall. In the cathode region, a complicated redistribution of the electric potential takes place forming a collisional double layer. The ionization rate (c) has a sharp peak in the negative glow and decreases sharply in the Faraday dark space before increasing again in the positive column. The ionization rate decreases near the anode forming an anode dark space and produces an off-axis peak on the anode surface.

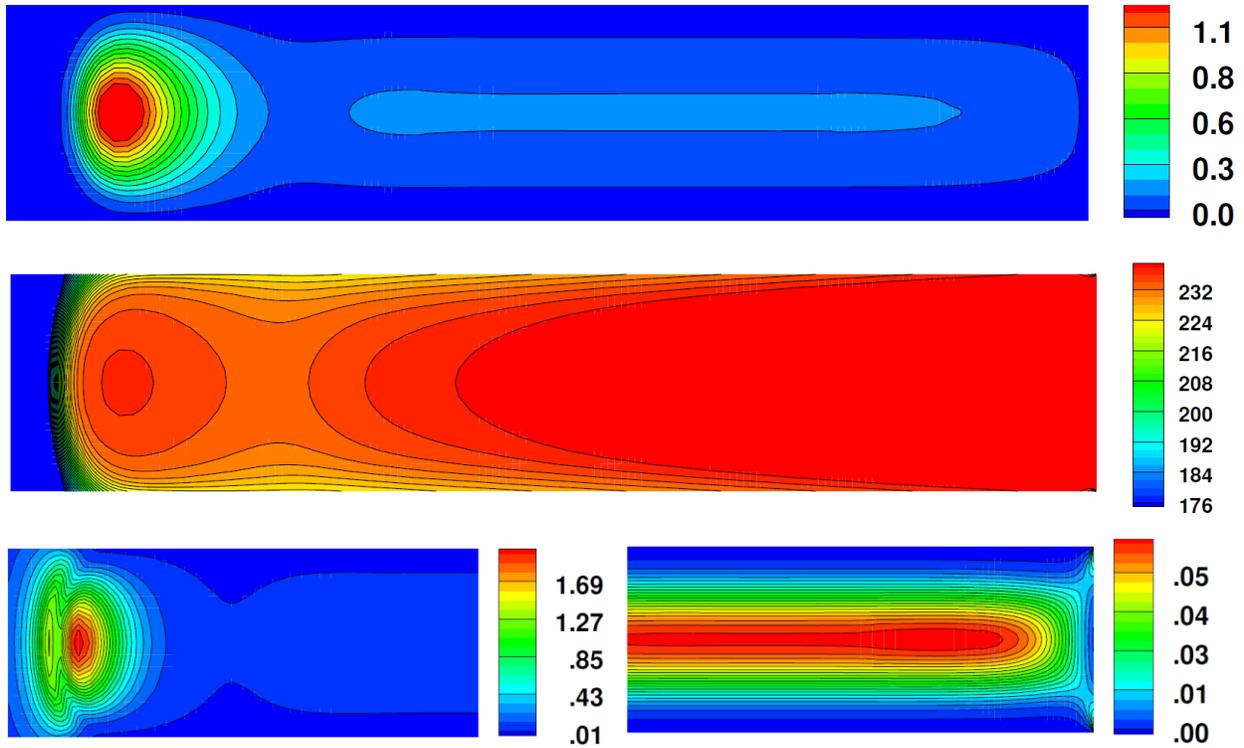

*Figure 3. Spatial distributions of electron density (a) in $10^{10}$ cm$^{-3}$, electrostatic potential (b) in V, and ionization rate (in $10^{15}$ cm/s) near cathode at and near anode (c,d).*

Figure 4 shows distributions of plasma parameters along the discharge axis. The ionization rate has a sharp peak near the boundary of the cathode sheath with plasma, it vanishes in the Faraday dark space, increases again in the positive column plasma, and decreases near the anode forming an anode dark space. The axial electric field changes its sign in two points *a* and *b* on the cathode region and remains constant in the positive column where the plasma density is axially uniform. The value of the axial electric field in the positive column is controlled by plasma to balance the ionization rate and the particle loss to the wall by diffusion and surface recombination. The field decreases near the anode and even reverses sign in the point *c* near the anode. The electron temperature has a large value in the cathode sheath, passes through a minimum in the cathode region before reaching a constant value in the positive column, and decreases near the anode. The calculated axial structure of the DC discharge contains all the key features observed in numerous experiments and described in textbooks on gas discharge physics.[46]

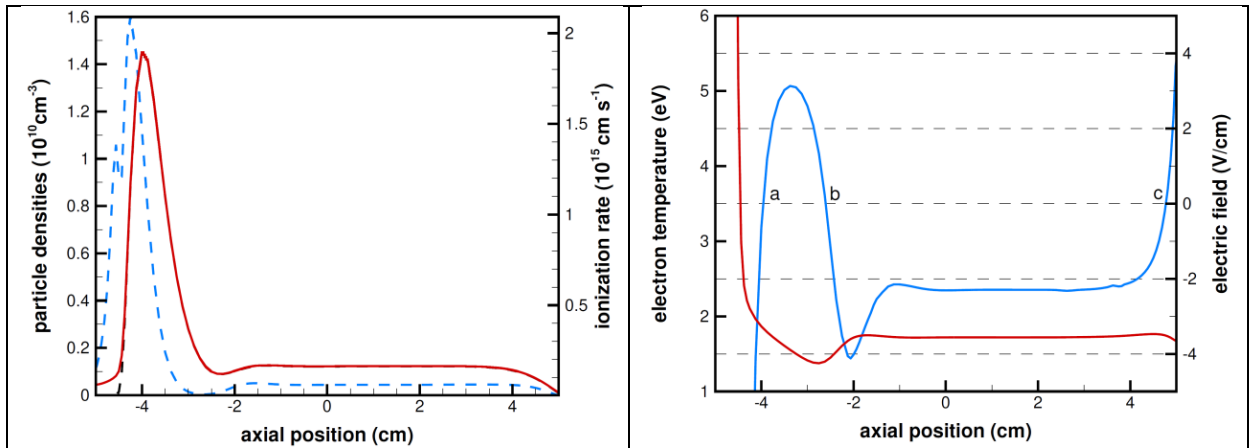

*Figure 4. Axial distributions of the electron (dashed) and ion (red line) densities and the ionization rate (dashed blue line) (a), electron temperature (red), and the axial electric field (b) on the axis.*

Figure 5 shows the radial distributions of the electron density and the ionization rate in the positive column and at the anode surface. In the positive column (a), the radial distributions of the ionization rate and plasma density are similar. At the anode, the radial distribution of the ionization rate forms a ring with a minimum on the axis and sharp peaks near the wall (b). At the same time, the radial distribution of the plasma density remains monotonic, as in the positive column. The formation of the anode ring that can break into separate spots along the azimuthal direction, have been observed in experiments and is of interest for understanding the self-organization of the anode region in DC discharges.

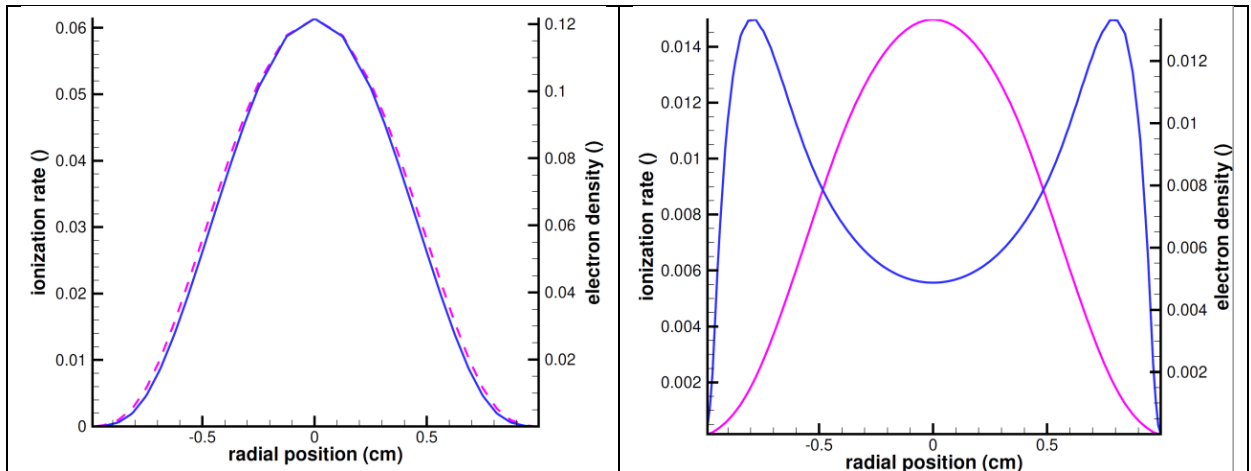

*Figure 5. Radial distributions of electron density (dashed pink line) and ionization rate (solid blue line) in the positive column (a), and at the anode (b).*

Figure 6 illustrates the structure of the cathode region based on 3D simulations of a DC discharge in a rectangular clamber with dimensions 1×1×3 cm for gas pressure 0.4 Torr and voltage 300 V. Iso-surfaces of the ion density as well as three slices with contour lines are shown in Figure 6. The conditions correspond to a normal discharge with the cathode sheath length of the order of the transverse chamber size $R$. The phenomenon of the normal current density is one of the key concepts in gas discharge physics.[46] The discharge is self-organized into a cathode spot, which covers only a part of the cathode surface. The current density in the spot remains constant, and the size of the spot increases with the increasing current until the spot covers the entire surface of the cathode. These processes have been previously described in numerous publications [47] and simulated with different codes.[48,49,50] It is seen in Figure 6 that the transverse distributions of the ion density gradually change from axially symmetric distributions on the cathode surface to a rectangular profile in quasi-neutral plasma.

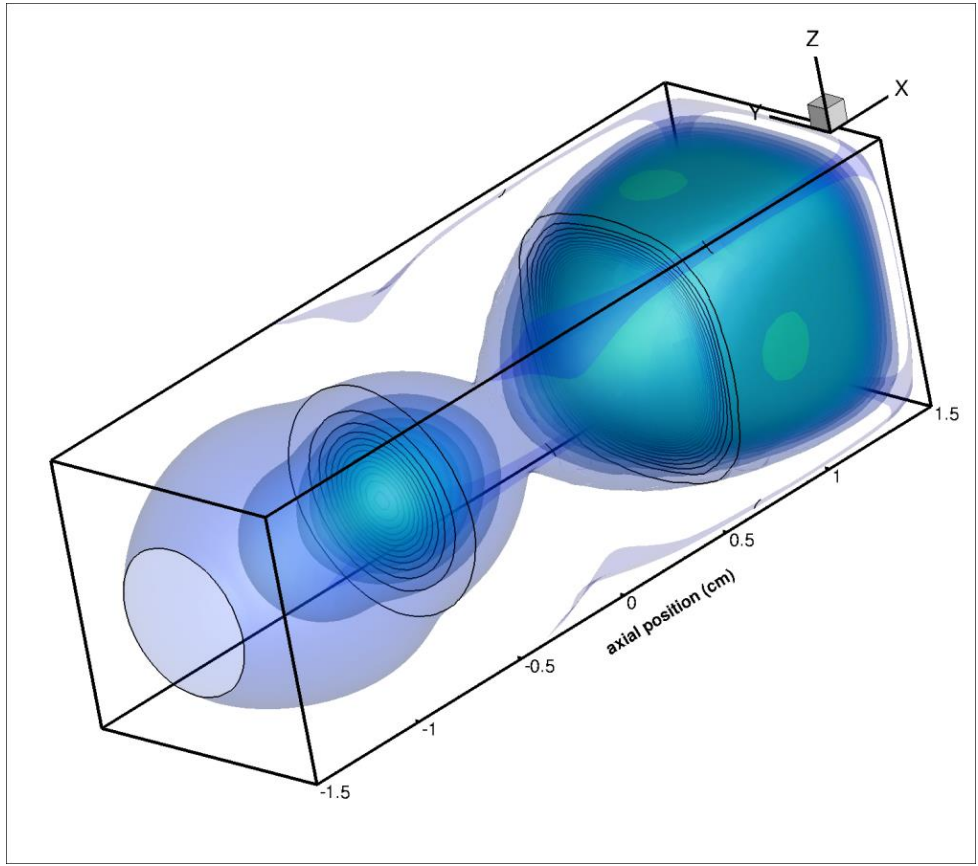

*Figure 6. Iso-surfaces of ion density with three slices with contour lines for a DC discharge in a rectangular chamber: Ar, 0.4 Torr, 2×2×6 cm, voltage 350 V.*

Figure 7 shows the axial distributions of the electron and ion densities and the electric field along the axis, which are similar to the 2D simulations in the cylindrical tube discussed above. The axial distribution of plasma density has two maxima: the first at the plasma-sheath boundary and the second close to anode. The electric field reverses sign three points marked *a, b*, and *c* in Figure 7. The first field reversal (point *a*) occurs close to the plasma-sheath boundary, the second reversal (point *b*) occurs near the minimum of plasma density, and the third field reversal (point *c*) is close to the point of maximum plasma density. The key difference with the previous 2D simulations is the lower current density in the 3D case and the short interelectrode gap, which corresponds to the absence of the positive column.

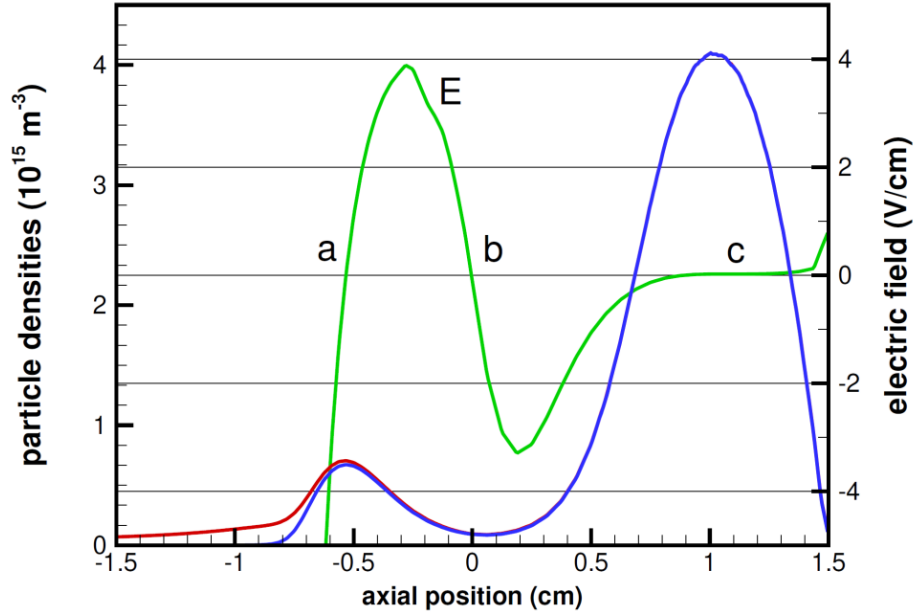

*Figure* 7. *Axial distributions of the electron (blue) and ion (red) densities and the electric field (green) on the axis for the conditions shown in Figure* 6. *The two electric field reversals at points* a *and* b *correspond to the formation of a collisional double layer in the cathode region.*

Our fluid model describes the typical structure of the cathode region observed in DC discharges and well captures qualitatively all the key features explained quantitatively by the kinetic theory.[51] The success of the fluid model in describing the overall structure of the cathode region is due to incorporating the energy balance of electrons into the model. Assuming that the ionization rate is a function of electron temperature rather than a local electric field allows capturing non-local effects caused by the high electron thermal conductivity. The new code has offered breakthrough capabilities for modeling DC discharges over a range of operating conditions.

### 4.2 Capacitively Coupled Plasmas

We have performed 2D axisymmetric simulations of CCP in a dielectric tube of radius $R = 6$ cm, inter-electrode gap $d = 2$ cm, driven by RF voltages at frequencies of 100, 10, and 1 kHz. The secondary emission coefficient was $\gamma = 0.1$. Figure 8 shows spatial distributions of plasma parameters in Argon at a pressure of 1 Torr, frequency of 100 kHz, and voltage 150 V. The time step in these simulations was selected between 2 and 10 ns.

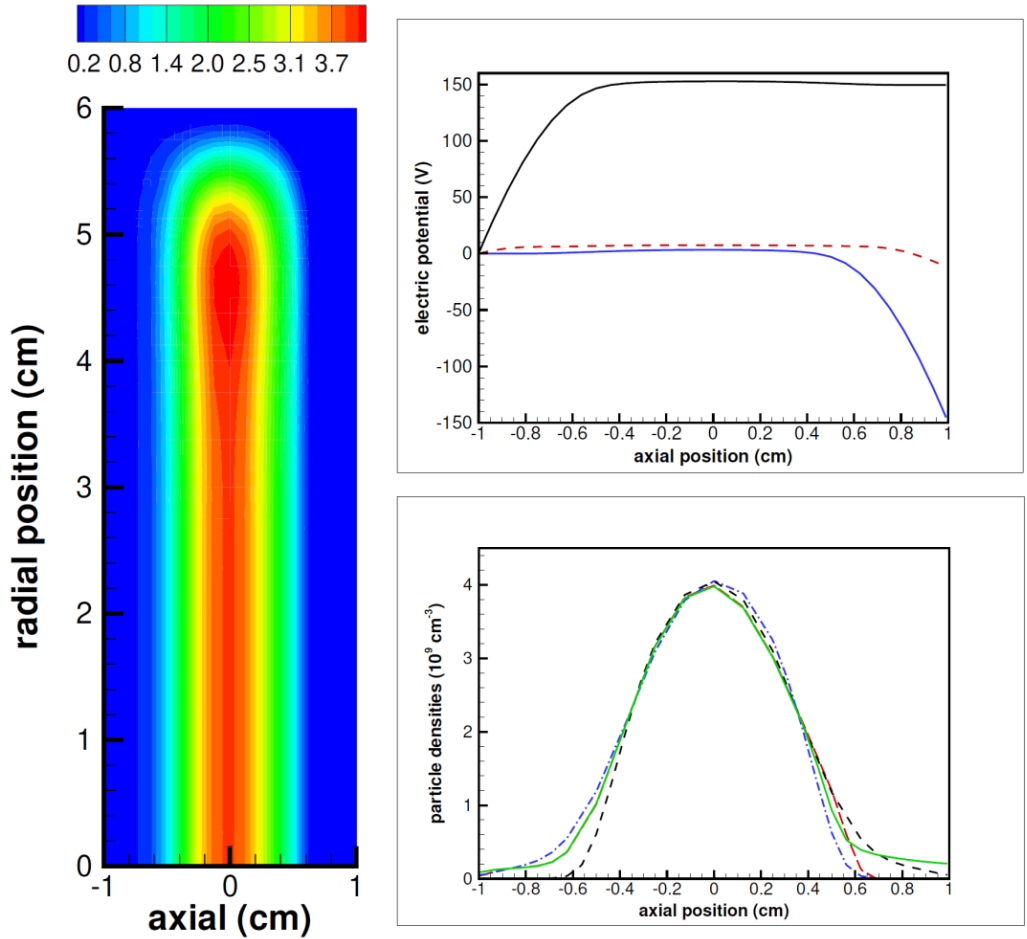

*Figure 8. Instantaneous spatial distribution of electron density (left) and axial distributions of the electric potential (top) and electron and ion densities (right) at three times during the RF period.*

Figure 9 compares the time modulation of electron density and temperature on the axis, in the center of discharge (at $x = 0$) for different driving frequencies. At 100kHz, the plasma density modulation in plasma is negligible, which is consistent with Figure 8. With decreasing driving frequency, the time modulations of electron density in plasma become substantial. The electron temperature is strongly modulated at all these frequencies.

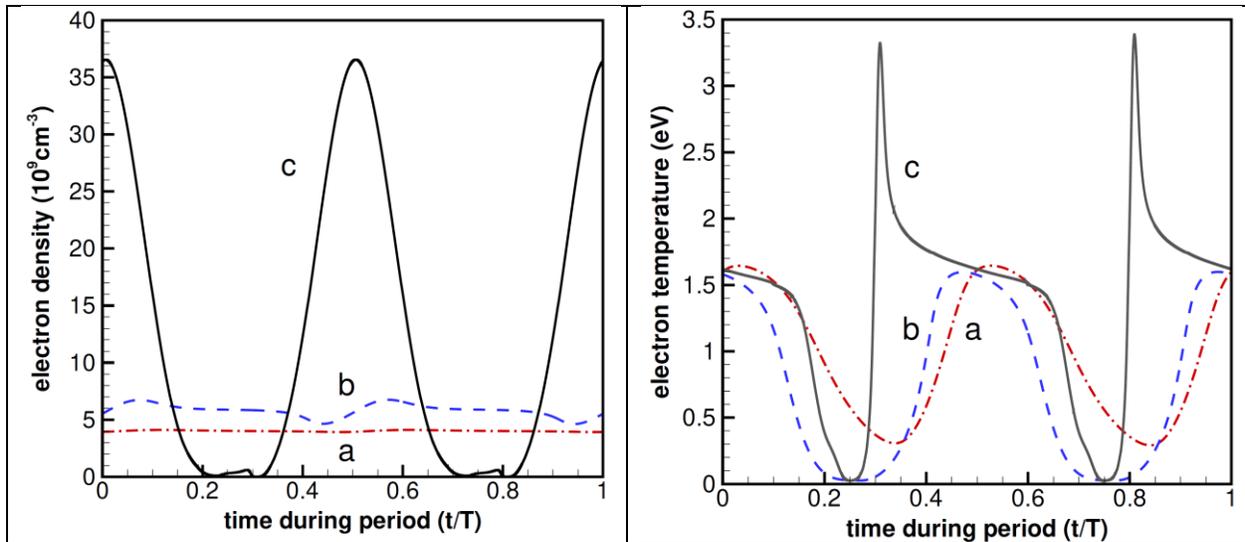

*Figure 9. Time modulation of electron density (left) and temperature (right) in plasma at frequencies 100 (a), 10 (b) and 1 (c) kHz.*

Figure 10 compares the time-variations of electron and ion current densities at the center of the electrode at 100 and 10 kHz. Both electron and ion currents are strongly modulated during the RF period. The negative values of the electron current are due to the secondary electron emission from the cathode, which reaches maximal value during the peaks of the ion current. The time modulation of the ion current is specific to low-frequency CCP operating in the so-called dynamics regime.[52] In this regime, the ion transit time through the sheath is comparable to the RF period, and the ion current at the electrode is absent during a substantial part of the RF period. At higher frequencies, the ions respond to the time-averaged value of the electric field in the sheath, the ion current at the electrode becomes nearly constant during the RF period, and the electron current has sharp peaks when the sheath voltage has minimal values and the ion current is constant during RF period.[53]

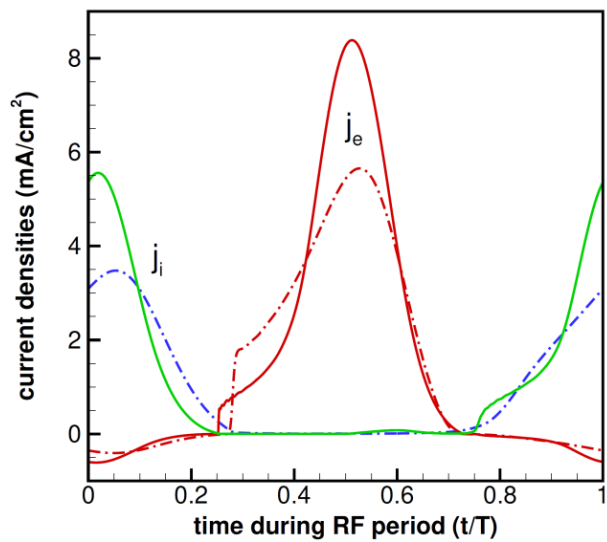

*Figure 10. Time modulation of the electron (red lines) and ion current densities (blue and green lines) at electrode for driving frequencies of 100 (dash-dot lines) and 10 (solid lines) kHz.*

## 4.3 Plasma Stratification in Noble Gases

Plasma stratification (i.e. pattern formation along the direction of discharge current) often occurs in DC and RF discharges over a wide range of gas pressures and discharge currents.[54,55] In DC discharges of noble gases, striations (ionization waves) usually move along the direction of the DC electric field, while standing striations are typical for molecular gases. We have already applied the new FNM plasma solver for simulations of striations in Argon discharges and briefly reported our results in two letters.[37,38] In these letters, we have discovered common nature of plasma stratification in DC and RF discharges in Argon gas at relatively high currents (plasma densities) where the nonlinear dependence of the ionization rate on electron density (described by Eq. (30)) was shown to be the main cause of stratification. Here, we provide further details of these simulations and also report some new results.

### Moving Striations in DC discharges

Simulations were performed for a cylindrical dielectric tube of radius $R = 1$ cm and length $L = 14$ cm. The anode was grounded, and a voltage was applied to the cathode. The electric potential at the dielectric walls was calculated from the local surface charge, which evolved in time based on electron and ion fluxes. During our simulations of the discharge development, striations first originate in the cathode region and gradually propagate towards the anode. However, they propagate towards the cathode that corresponds to the backward waves (with the group and phase velocities in opposite directions), which are well known from the analytical theory of small-amplitude striations under these conditions. The corresponding movie illustrating the dynamics of discharge stratification and mesh adaptation is available on the journal web site.

Figure 11 shows instantaneous spatial distributions of electron density in diffuse and constricted discharges at two gas pressures (1 Torr and 400 Torr) in Argon. The radius of the plasma column in the diffuse discharge is controlled by surface recombination, and the striation wavelength is about the radius of the discharge tube. In the constricted discharge, the radius of the plasma column is controlled by volume recombination and is smaller than $R$. The striation wavelength is smaller, and the plasma radius changes over striation wavelength during the wave propagation. This agrees well with the reported experimental observations of the striations in constricted discharges. The bottom part of Figure 11 shows a dynamically adapted mesh for the constricted discharge. The mesh adaptation criterion is based on the gradients of the electrostatic potential and electron density. A movie available at the journal web site illustrates the dynamics of plasma stratification in DC discharges. Striations originate from the cathode region and propagate towards the anode. However, they move from anode to cathode, which corresponds to backward waves observed in experiments.

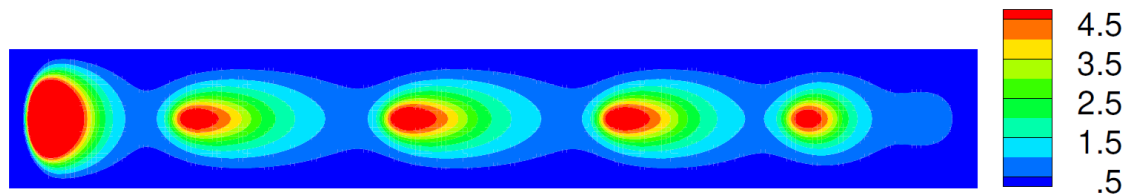

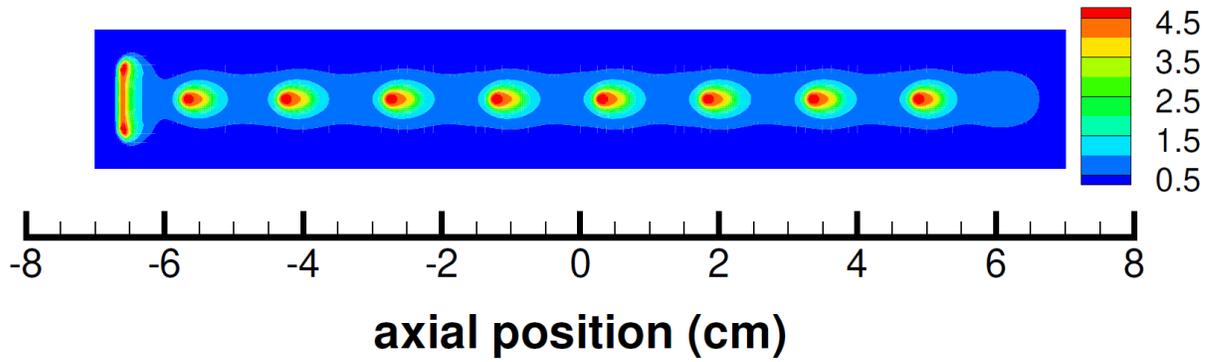

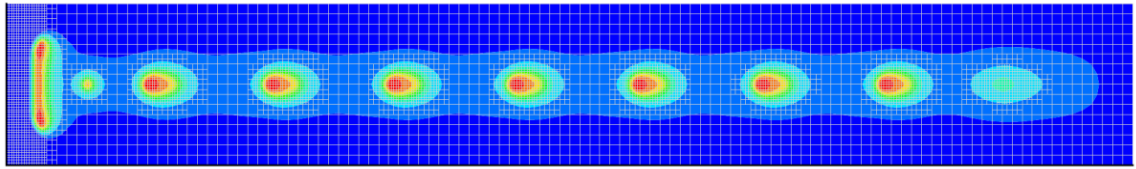

*Figure 11. Instantaneous spatial distributions of electron density ($10^{10}$ cm$^{-3}$) for the diffuse discharge (top) and constricted discharge (middle). The bottom part shows the dynamically adapted mesh for the constricted discharge (Argon 400 Torr).*

Figure 12 shows distributions of the electric field, electron density, and temperature on the axis of the diffuse (left) and constricted (right) discharges. In the diffuse discharge, the maximal value of the electric field is observed near the maximal gradient of plasma density, which corresponds to the dominance of the ambipolar component of the electric field over the conduction component. The maximal value of the electron temperature is shifted towards the cathode compared to plasma density, which corresponds to the propagation of striations towards the cathode. Electric field reversals take place in the constricted discharge, which indicates that striations are substantially non-linear in the constricted positive column. In diffuse discharge, electric field reversals are negligible for these conditions.

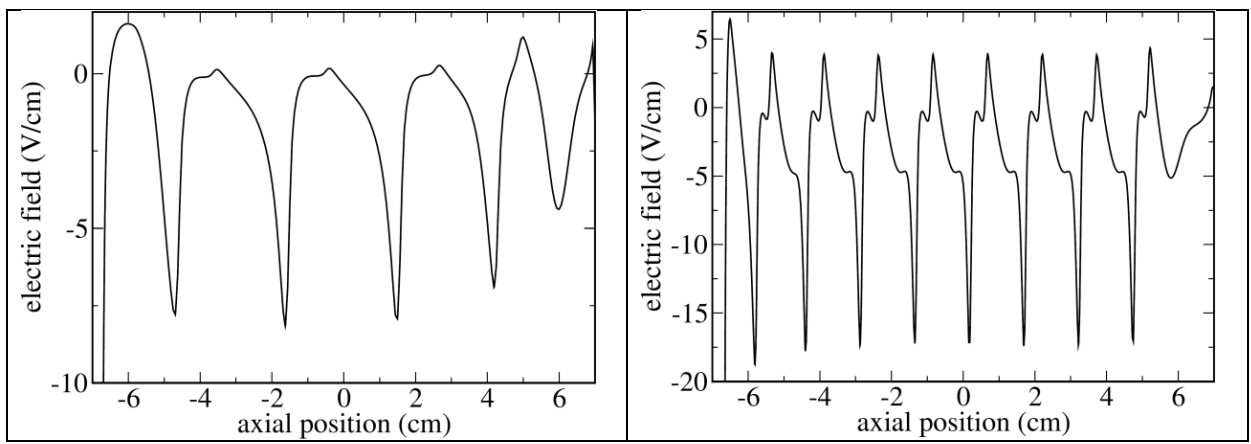

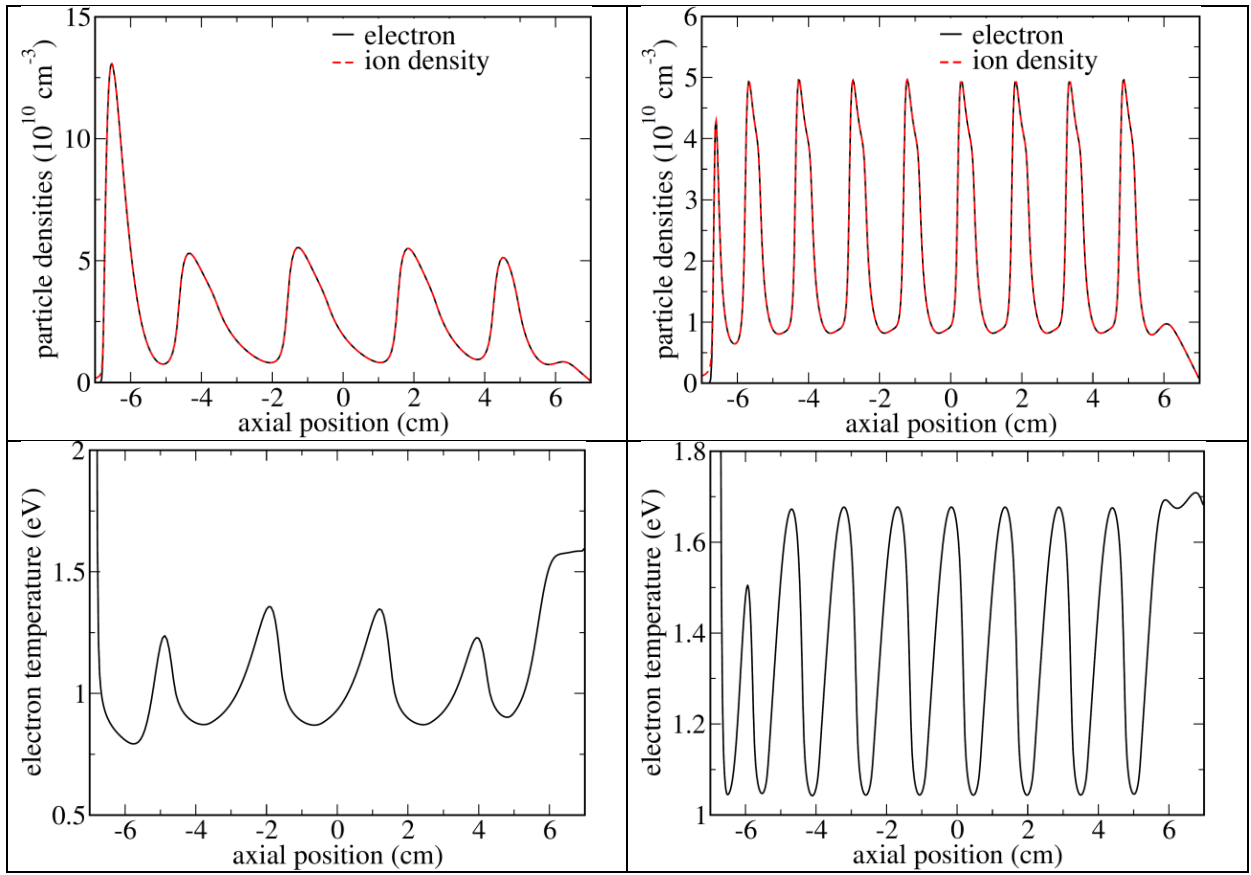

*Figure 12*: *The axial distributions of the normalized electric field, electron density, and temperature in the diffuse (left) and constricted (right) discharges.*

In the constricted discharge, the maximal value of the electric field is observed again near the point of maximal plasma gradient of plasma density. This corresponds to the dominance of the ambipolar electric field over the conduction field. The field is close to zero at the points of maximal plasma density on the axis. That is in good agreement with the two-dimensional theory of striations in constricted discharges. The maximal values of the electron temperature are shifted towards the cathode compared to the maximums of plasma density, which corresponds to the propagation of striations towards the cathode. The electric field changes sign between the maximum of plasma density and the maximum electron temperature. This corresponds to highly non-linear waves under these conditions.

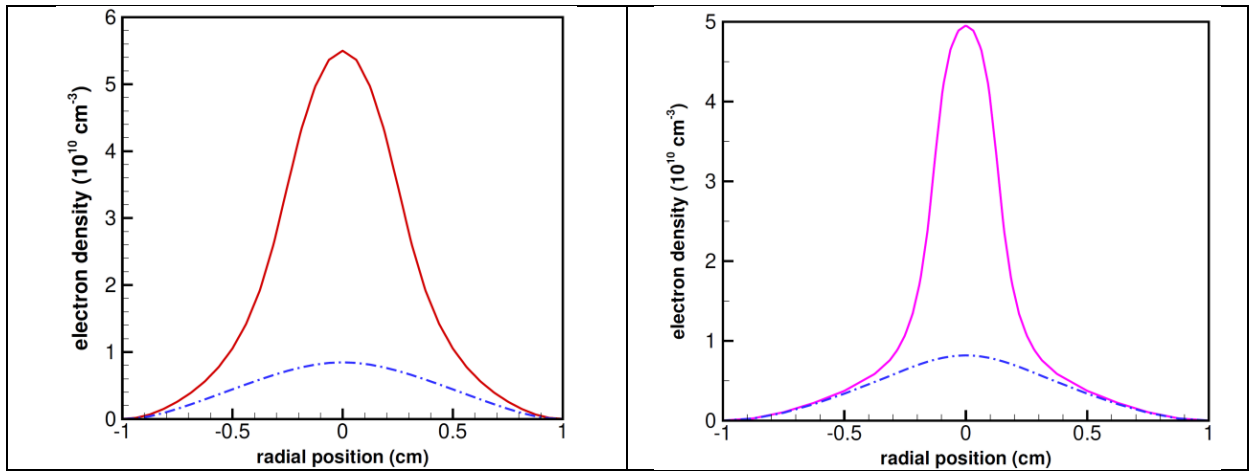

*Figure 13: Radial distributions of electron densities in two phases of striation corresponding to the maximal and minimal density on the axis in the diffuse (left) and constricted (right) plasma.*

Figure 13 shows the radial distributions of electron densities in two phases of waves, which correspond to the maximal and minimal density on the axis. In the diffuse discharge (left), the radial distribution of plasma densities changes weakly. In the constricted discharge (right), the radial distribution of plasma density changes substantially over the striation wavelength. The radius of plasma has a minimum at the point of maximal plasma density on the axis, which is in good agreement with the two-dimensional theory of striations in constricted discharges.

### Standing Striations in RF discharges

The key difference between moving striations in DC discharges and standing striations in RF discharges is the absence of the time-average component of the RF electric field in the striation-free positive column. We have observed in our experiments and simulations that any discharge asymmetry and the appearance of even small DC component of the electric field in CCP results in a movement of striations along the axis. Figure 14 shows an example of 2D axisymmetric simulations of CCP for tube radius R = 1 cm, inter-electrode lengths L = 14 cm (a) and 20 cm (b), gas pressure 0.5 Torr, driving frequency of 20 MHz, and $n_c$ = 4×10$^{15}$ m$^{-3}$. The striation wavelength varies discontinuously with changing the inter-electrode distance in such a way that an integer number of standing waves always forms between the electrodes. Six striations are observed in our simulations for L= 14 cm, and 8 striations are formed for L=20 cm. It is seen that the two central striations at L=20 cm are slightly longer than the rest of the waves. They will break into an additional pair with a further increase of the inter-electrode distance L.

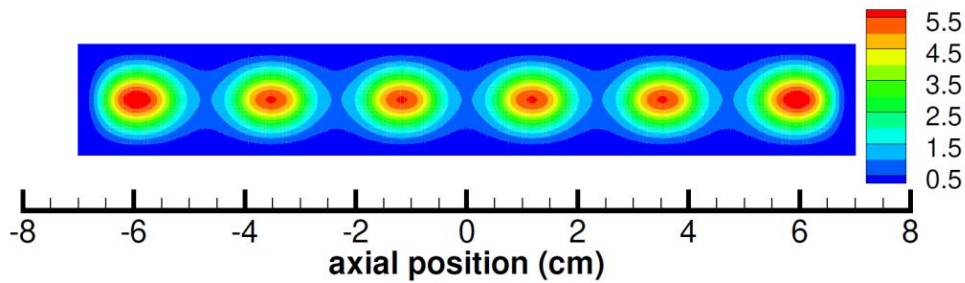

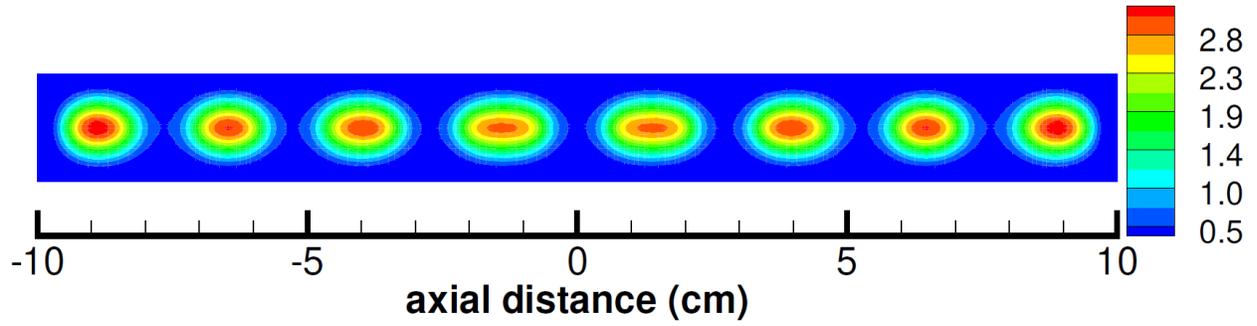

*Figure 14. Spatial distributions of the electron density ($10^{10}$ cm$^{-3}$) obtained in 2D-axisymmetric simulations of CCP at 20 MHz for L= 14 cm (a) and L=20 cm (b)*

Figure 15 shows the time variations of electron density in the middle of the gap (at $x = 0$) and close to the electrode (at $x = L/2$) during the plasma stratification process. It is seen that standing striations are formed during a few thousands of RF cycles. The time of striation formation is also large compared to the particle ambipolar diffusion time to the wall. A movie available on the journal web site illustrates the transient process of the striation formation. They propagate from electrodes towards the discharge center. As discussed in Ref. [38], in fully striated plasma, despite substantial time variations of the power deposition, not only the electron density but also the electron temperature does not vary in time under these conditions, except within the oscillating sheaths near the electrodes.

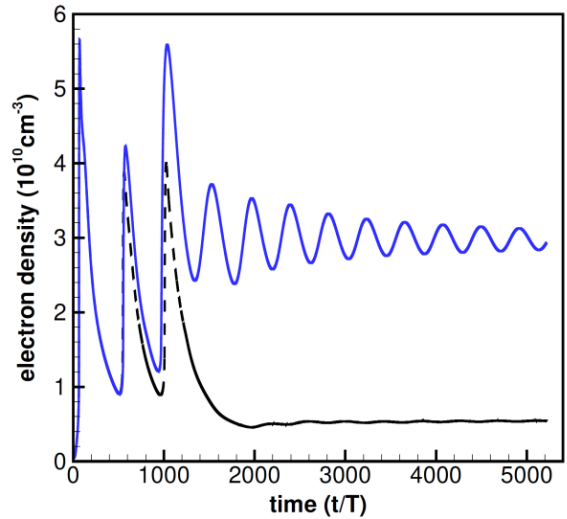

*Figure 15. Oscillations of plasma density on the axis at x=0 (dashed black line) and x=L/4 (solid blue line) obtained in 2D axisymmetric simulations of a 20 MHz CCP with L= 14 cm.*

## Impact of electron thermo-diffusion on plasma stratification

As discussed earlier, the implemented FNM treats the thermo-diffusion in the electron energy equations as a separate term in Eq. (9). This allows one to use different models for the related coefficient (e.g., expressed via different $c_e$ coefficients). Such a treatment is not possible using the conventional plasma modeling approach in Eq. (11) since it is usually written in an SG form which does not allow independent control of the thermal diffusion coefficient. In our formulation taken from the semiconductor models, it is easy to study the impact of the thermo-diffusion coefficient on the predicted standing striations in RF discharges. Figure 16 shows that for the nominal thermo-diffusion coefficient, six standing striations are formed. By doubling the thermo-diffusion coefficient, we obtain only four striations. Generally, with decreasing the thermo-diffusion coefficient, the number of striations increases while the amplitude of striations decreases. Striations eventually disappear in the limit of weak thermal diffusion. This indicates that the thermal diffusion of electrons plays a major role in selecting the optimal wavelength of the striations observed in the experiments. The developed code helps to clarify the plasma stratification mechanism not only at the linear stage but also at the non-linear stage often observed in the experiments.[38]

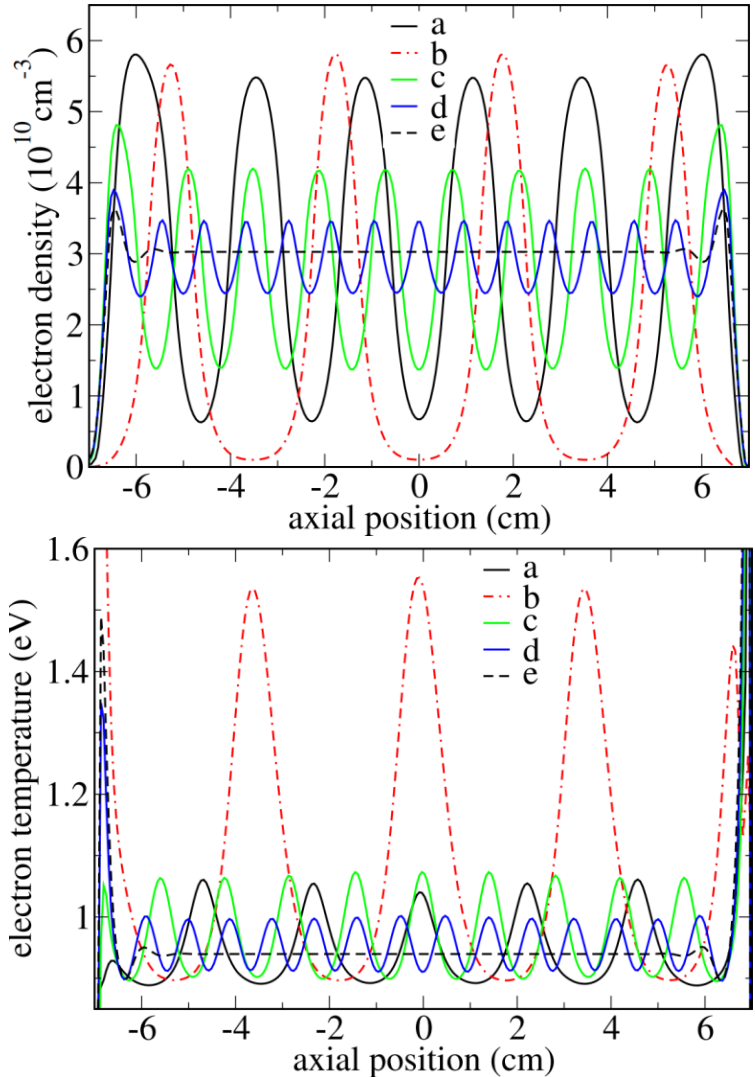

*Figure 16. The axial profiles of electron density (top) and (instantaneous) electron temperature (bottom) for 20 MHz CCP: a) normal thermodiffusion coefficient ($\kappa_e$), b) $\kappa_e \times 2$, c) $\kappa_e/8$, d) $\kappa_e/40$, e) $\kappa_e/80$.*

### 4.4 Comparison of Explicit and FNM Solvers

Detailed benchmarking of the previously developed explicit code and the FNM code was not the goal of the present work and can be a subject of a separate study. In this section, we provide some comments and estimates to compare the accuracy and efficiency of the two codes. The programming language for both codes is C with inheritance.[31] Both the explicit and FNM codes were run on Intel Xeon CPU-E5-2680 v4

2.40 GHz processors in serial mode. However, the numerical schemes in the explicit code [30] and the new FNM code are quite different. In particular, the explicit code used the semi-implicit Poisson solver, fully explicit isothermal SG scheme for the particle transport, and a scheme with no special treatment of the Joule heating term and boundary conditions. Hence the results of the two codes can be expected to be somewhat different even on the same computation grids. Direct benchmarking of the explicit and FNM codes would require the implementation of the same spatial and temporal discretization schemes, with the same treatment of different terms of the governing equations.

However, some estimates can still be made using the two codes in terms of the memory footprint and timings for comparable accuracy of the solution. We have selected two 2D axisymmetric, steady-state problems for comparative studies. One problem is a short DC plasma case (400 mTorr of Ar gas, gap 2 cm, applied voltage 200 V) and the other problem is a long DC plasma case (1 Torr of Ar gas, inter-electrode distance 5 cm, tube radius 1 cm, applied voltage 100 V). The number of solved species was 5 in both cases: electrons, ground-state Ar species (allocated but not solved for), and (for generality) 3 types of ions, of which only the $Ar^+$ ion played a role under simulated discharge conditions. The short DC plasma case has a distinctive feature of stronger electric fields between the electrodes, and thus shorter CFL time steps allowed in the explicit code. The second case is characterized by larger plasma formation times due to the presence of a long positive column region controlled by the ambipolar diffusion.

For each setup, we have selected the same AMR grids (with cell clustering around the cathode, anode, and dielectric walls regions) to be used in the explicit and FNM codes. The number of cells was close to 2K in both cases. RAM required to run the explicit code was ~350 MB while the FNM code required ~400 MB (~0.3-0.4MB per cell). The extra RAM of ~50 MB required to run the FNM code could be attributed to the Jacobian matrix storage, as well to the linear matrix solver storage (which was found to scale exceptionally well with the number of degrees of freedom, see references in the paper). The explicit code relied on the semi-implicit implementation of the Poisson solver [30] while the Poisson equation itself was solved by the multigrid technique [31] with the convergence of 3-4 orders of magnitude (typically, in 4-5 iterations) thus ensuring implicitness of this solver. The results showed that 1000 time steps for the explicit code took ~25 s of CPU time, while 1000 steps of the implicit code took ~500 sec, thus the explicit code being ~20 faster per time step than the FNM code. The physical time steps for the explicit code were about 10-50 ps at the initial stage of plasma evolution and later dropped to several ps after the narrow plasma sheaths were formed.

On the contrary, the physical time step in the FNM code started with ~100 ps and could be gradually ramped to 0.1 µs (sometimes even larger, up to 1 µs) during the first several 1000's steps, which then remained at this level until the end of the simulation. We have run these simulations to a physical time of 1 ms, which is a typical time scale for the plasma to establish due to chemical reactions and ambipolar diffusion processes. For the explicit code with a time step of ~10 ps to reach out to such times, it requires ~100M (1 ms / 10 ps) time steps. For the implicit code with the time step of 0.1 µs, it requires only ~10K time steps. While the implicit code was found to be ~20 times more expensive, because of the drastic difference (~10,000X) of the time steps required, the final simulation could be obtained ~500 times faster in terms of the CPU time. Since the plasma can be established faster than 1 ms (say, 0.2 ms), the resulting (conservative) speed up estimation is 100X.

It is worth noting that our explicit code has the semi-implicit Poisson solver thereby avoiding the strict dielectric relaxation time step limitation (which can be a fraction of ps at the studied plasma densities); hence, the explicit time step was only limited by the CFL condition. The FNM solver relied on the explicit chemistry implementation, which was likely to be the main cause of its observed time step limitation. More accurate benchmarking of the explicit and FNM codes could involve a larger set of cases (e.g., with a larger number of cells and plasma species, as well as time transient cases, such as RF plasmas). Such benchmarking could be a subject of separate studies. However, even these preliminary results have already shown the great advantages of developed FNM technology. This technology is currently being further advanced via fast-slow modular splitting and chemistry implicitness for efficient plasma simulations.

## 5 Conclusions and Outlook

We have described an initial implementation of a multi-fluid, multi-temperature plasma solver with adaptive Cartesian mesh using a full-Newton (coupled, implicit) scheme. The new solver provided breakthrough capabilities for solving several gas discharge problems that we could not solve before with existing software. With the new solver, we could calculate the two-dimensional structure of the entire DC discharges including cathode and anode regions with electric field reversals, a normal cathode spot, and an anode ring, for realistic values of the secondary electron emission coefficient at the cathode and high discharge currents. We could also simulate moving striations in diffuse and constricted DC discharges and standing striations in RF discharges in Argon gas. We have observed good convergence rates and proper dynamic grid adaptation for resolving time-variations of space-charge sheaths in CCP at the electron scale, as well as the slow plasma dynamics at the ion time scale. The fully implicit treatment of the coupled plasma equations allowed using large time steps (speedup factors up to 100 compared to explicit solvers), and the full-Newton treatment enabled fast non-linear convergence at each time step, offering greatly improved efficiency of multi-fluid plasma simulations.

In our initial implementation presented in this paper, the plasma boundaries were aligned with the grid lines. The treatment of non-aligned boundaries is planned for future work using the cut-cell technique, as was previously done in our explicit plasma solver. The ability to embed solid objects of complex shape defined analytically or via CAD-generated surfaces[31,56] appears very attractive for modeling plasma systems in multi-dimensional settings, especially for modeling discharges with liquid electrodes[11] using the VoF method for tracking free material interfaces with AMR. It is expected that such problems can be efficiently treated by the developed FNM-ACM technique. We expect that the parallelization of the FNM-ACM solver can be done efficiently using the Forrest-of-Trees or space-filling curve algorithms. The work in these directions is currently underway and will be reported elsewhere.

Adding a general-purpose chemistry solver is another subject of our future work. The developed solver is ready to be linked with finite-rate chemistry solvers using multi-temperature reaction rates (for electrons, gas species, and vibrational states of molecules). The non-linear source and loss terms for charged and neutral species can be treated using several techniques, such as a point-implicit approach,[57] high-order Runge-Kutta techniques with adaptive time-stepping, as well as explicit techniques. The most general technique is to treat the species transport and finite-rate chemistry in a fully coupled way, which is commonly used in CFD for modeling combustion and flames with complex and often stiff chemistries. In CFD, stiffness of the governing system of equations comes, among other things, from the necessity to accurately resolve the thermal and kinetic boundary layers (which require dense meshing), as well as from the chemical reactions which are often strongly nonlinear and contain the time scales that greatly differ from those of the convective and diffusive terms. A similar challenge in discharge plasmas is ionization in (usually thin) sheaths regions and the plasma bulk often evolving on vastly different time scales. The need for coupling the charged particle motion and the electron energy transport with self-consistent electric fields makes plasma processes highly non-linear and multi-scale problems. The developed FNM-ACM technique is expected to offer many benefits for tackling the disparity of the time scales and non-linearity of plasma systems.

We wish to point out that the implemented code has not been fully optimized for maximal performance. However, the code runs fast and allows using large time steps for stability with good convergence observed for a broad range of variation in the grid resolution. The time steps employed in the present paper can be further increased by implicit (or semi-implicit) treatment of the finite-rate chemistry, as was discussed earlier. Among further optimization techniques that can be implemented in the future is an automatic time-step-control algorithm,[18] which ensures that maximum allowable time steps are employed during runs. Another technique is the matrix bandwidth reduction, which can further speed up the code. Such a technique is easily realizable with ACM by reordering the cells for decreasing the adjacency matrix. Another advantage of the FNM is that it allows efficiently treating the cut cells that often invoke the small-cell problem, which can severely limit the allowed time step in explicit algorithms and is usually treated by rather complicated (in terms of bookkeeping) cell-merging techniques.[31] The FNM technique allows avoiding such complicated techniques all together by including the cut cells into a single computational matrix along with the off-body cells. Since the particle-field coupling challenge can become even stronger in the cut cells than in the bulk plasma cells, the use of implicit (thus, no CFL limit) but uncoupled schemes can result in severe convergence problems, especially in small cut cells with volumes much smaller than off-body cells. This is where the full advantage of the developed FNM technique can become even more relevant.

For multi-fluid modeling of collisional gas discharge plasmas, we have adapted the numerical techniques extensively used in the state-of-the-art, commercial semiconductor modeling simulators. In these simulators, fully coupled techniques are used for electrons and holes in a single (global) coupled matrix. For modeling plasma systems, in general, this approach may not be necessary, as the electron and ion mobilities in plasmas differ by factors of 100−1000 whereas the charge carrier mobilities in semiconductor systems may be of the same order of magnitude, depending on the conduction and valence band physics. Then, for the plasma systems, ion transport can be solved separately, which can give a significant reduction of the computational cost, especially for modeling of more exotic plasma systems such as ion-ion plasmas, strongly magnetized plasmas, or "liquid" plasmas. The developed FNM approach can be combined with the time-scale separation methodology to treat electrons and heavy (ion and neutral) plasma species by loosely dependent modules (as was done in Ref [5]). The loosely coupled approach will allow different treatment for electrons and ions, with electrons treated by hybrid kinetic-fluid models, while ions treated by multi-fluid models. Furthermore, such an approach will facilitate solving problems with large numbers of heavy species and complicated chemical mechanisms.

## Acknowledgments

This work was partially supported by the NSF EPSCoR project OIA-1655280 "Connecting the Plasma Universe to Plasma Technology in AL: The Science and Technology of Low-Temperature Plasma".